\documentclass[nofootinbib,aps]{revtex4}
\usepackage[dvips]{graphics}
\usepackage{amsfonts}
\usepackage{graphicx}

\newcommand{\Msat}{{{M_{\rm sat}}}}
\newcommand{\dedm}{{{\frac{\partial e}{\partial m}}}}

\begin{document}
\bibliographystyle{prsty}

\title{Frustrated quantum magnets}

\author{Claire Lhuillier}
\email{lhuilier@lptl.jussieu.fr}
\affiliation{Laboratoire  de    Physique Th\'eorique des Liquides
Universit\'e P. et M. Curie  and UMR 7600 of CNRS
Case 121, 4 Place Jussieu, F-75252 Paris Cedex}

\author{Gr\'egoire   Misguich}
\email{misguich@spht.saclay.cea.fr}
\affiliation{Service de Physique Th\'eorique,
CEA Saclay, F-91191 Gif-sur-Yvette Cedex}
\date{Sept. 7th, 2001}

\maketitle


In these lectures~\footnote{Lectures  presented at the  Cargese summer
school  on ``Trends in high-magnetic field  science'',  May 2001.}  we
sketch a rapid survey  of recent theoretical advances  in the study of
frustrated quantum magnets with a  special emphasis on two dimensional
magnets.  One dimensional problems   are only very  briefly discussed:
this field has been extraordinarily flourishing during the past twenty
years both   experimentally   and  theoretically.  In   contrast   the
understanding of two dimensional quantum magnets is much more limited.
The number of unexplained experimental  results is probably  extremely
large, unfortunately the theoretical tools to deal with exotic quantum
phases in 2 dimensions  are still rather limited.  In the absence of a
significant amount   of exact  results,   the  picture drawn  in   the
following sections is based on the comparison of various approaches to
$SU(2)$ spins: series   expansions,  exact  diagonalizations,  Quantum
Monte-Carlo    calculations  and    the   hints    got   from  large-N
generalizations.

From these comparisons we suggest  that  one could  expect at least  4
kinds of different low temperature physics in two-dimensional systems:
\begin{itemize}
\item semi-classical N\'eel like phases
\item and three kinds of purely quantum phases.
\end{itemize}
These 4 phases are the subject of the  first 4 sections of this paper:
their properties are  summarized in Table~\ref{4phases}.   The quantum
phases appear in situations where  there are competing interactions, a
high degree of frustration  and a rather low  coordination number.  In
section
\ref{section_mag} 
we discuss the very rich magnetic phase  diagram of frustrated quantum
magnets, which can exhibit   hysteretic metamagnetic transitions   and
magnetic plateaus at zero and at rational values of the magnetization.
Chiral phases,   and   the  possible  relation  between  magnetization
plateaus and Hall conductance   plateaus specifically studied  in some
other lectures of this School are briefly discussed.

\begin{table}
\begin{center}
\begin{tabular}{|c||c|c|c|c|c|}
\hline
Phases& G.-S. Symmetry Breaking&Order Parameter	&First excitations\\
\hline
\hline
&SU(2)&& \\
 Semi-class. N\'eel order & Space Group&Staggered Magnet.& Gapless magnons\\
& Time Reversal&& \\
\hline
& &dimer-dimer LRO& all excitations are gapped\\
Valence Bond Crystal&Space Group&or S=0 plaquettes &Confined spinons\\
&&LRO&thermally activated $C_v$ and $\chi$\\
\hline
 && No local &all excitations are gapped\\
 R.V.B. Spin Liquid&topological degeneracy&order parameter& Deconfined spinons\\
(Type I)&&&thermally activated $C_v$ and $\chi$\\
\hline
&&& Singlet excitations are gapless\\
	&&No local&Triplet excitations are gapped	\\
 R.V.B. Spin Liquid&topological degeneracy	&order parameter& Deconfined spinons\\
		&	&&  T=0 entropy\\
(Type II)&&&thermally activated $\chi$\\
&&&  $ C_v$   insensitive to magn. field at
low T\\ 
\hline

\end{tabular}
\end{center} 
\caption[99]{The four 2-dimensional phases described in the four
first sections.}
\label{4phases}
\end{table}

To complete Table~\ref{4phases} let us underline that N\'eel
ordered magnets are characterized by a unique energy scale
(essentially given by its Curie-Weiss temperature $\theta_{CW}$,
 which is
directly related to the coupling constant of the Hamiltonian).
All the quantum phases studied up to now display a second energy
scale which can be an order or two order of magnitude lower than 
$\theta_{CW}$ and mainly associated with a spin gap. It is the
range of temperatures and energies under study in this
lectures.

\begin{section}{Semi-classical ground-states}
\subsection{Heisenberg problem}
In most of the cases the ground-state of the
antiferromagnetic spin-1/2 Heisenberg Hamiltonian
\begin{equation}
{\cal H} = 2 J \sum_{<ij>}  {\bf S}_i\cdot {\bf S}_j,
\label{heis}
\end{equation}
with $J$  $>0$ and the sum limited to first neighbors,
is N\'eel ordered at $T=0$. 
In 3 dimensions, when increasing the temperature the N\'eel
ordered phase gives place to a paramagnet through a 2nd
order phase transition $( T_N \sim {\cal O } (J))$. In 2
dimensions the ordered phase only exists at $T=0$ (Mermin
Wagner theorem
\cite{mm66} 
). In one dimension, even at $T=0$, there is
only quasi-long range order with algebraically decreasing
correlations (for an introduction to spin systems see~\cite{m81,c89a,auer94}).

N\'eel ground-states break the continuous $SU(2)$ symmetry and support
low lying excitations  which are the  Goldstone  modes associated with
this broken symmetry.  These excitations, called magnons,  are $\Delta
S^z =1$ bosons that can be pictured as long  wave-length twists of the
order parameter  (the  staggered magnetization).   They form  isolated
branches of excitations described by their dispersion relation $\omega
({\bf k})$.  $\omega  ({\bf k})$ vanishes  in reciprocal space  at the
set  of {\bf k}  vectors $ \{\tilde {\bf k}  \}$ characteristic of the
long range order: {\it i.e.} the wave vector ${\bf k=0}$ and the peaks
in the structure function
\begin{equation}
{\cal S}( {\bf q}) =\int d^{D}{\bf r}\;e^{i{\bf q}\cdot{\bf r}}   {\bf S}_0\cdot {\bf S}_{\bf r},
\label{structurefonction}
\end{equation}
where D is the lattice dimensionality.
 
The  geometry   of  the N\'eel   order parameter   can  in  general be
determined through a classical minimization  of Eq.~\ref{heis}.  It is
a two-sublattice up-down  order (noted $ud$  in the  following) in the
case of the square  lattice, a three-sublattice with  magnetization at
$120$ degrees from each other in the triangular case.

This   classical approach   neglects   the quantum   fluctuations: the
classical solution $ud$ is an eigenstate of  the Ising Hamiltonian but
not of  the  Heisenberg  Hamiltonian, the extra    $X-Y$ terms of  the
Heisenberg Hamiltonian induce pairs of spin-flips and reduce the value
of  the staggered   magnetization.  For  a   moderate reduction,   the
spin-wave calculation  is  usually a  good approximation  to take this
quantum  effect into    account~\cite{a52}.  Introduction  of  quantum
fluctuations
\begin{itemize}
\item
renormalizes the ground-state energy through  the zero point
fluctuations
\item decreases the sublattice magnetization by a
contribution
\end{itemize}
\begin{equation}
\delta \propto  \int {\frac {g(k)} {\omega (k)}} d^{D}k
\label{reduc}
\end{equation}
where $\omega (k)$ is the dispersion law and
$g(k)$ is a smooth function of k, depending on the lattice,
which is non zero in the whole Brillouin zone.
It is a straightforward
exercise to show that $ \omega (k) \propto k$ for small k.
Eq.~\ref{reduc}  indicates the  absence of  long  range  order (LRO) in
$D=1$  dimension through the breakdown  of the spin-wave approximation
(divergence of the integral).

The reduction  of the order  parameter with  respect  to the classical
saturation value   $M/M_{cl}$ increases  when the  coordination number
decreases; it is larger on triangular based lattices than on bipartite
ones (see   Table~\ref{ener-_param}). This last  feature  is sometimes
called ``geometrical frustration''. In the framework of the Heisenberg
model  this   should  be understood   in    the following sense:   the
ground-state of the Heisenberg   problem on the triangular  lattice is
less  stable than the ground-state of  the  same problem on the square
lattice (and bipartite lattices), both in the classical regime ( where
$ {\left<{\bf    S}_i\cdot  {\bf  S}_j\right>}^{cl}_{sq}=-0.25$    and
${\left<{\bf  S}_i\cdot {\bf S}_j\right>}^{cl}_{tri}=-0.125$), as well
as in  the $SU(2)$    quantum  regime ($  {\left<{\bf   S}_i\cdot {\bf
S}_j\right>}^{qu}_{sq}=-0.3346$,  whereas $ {\left<{\bf S}_i\cdot {\bf
S}_j\right>}^{qu}_{tri}=-0.1796$).

 The  ground-state of the  spin 1/2 Heisenberg  model (Eq. 1) does not
 have N\'eel order on the following lattices:
\begin{itemize}
\item the Bethe Chain (exact result)
\item the kagom\'e lattice (at the classical level there is
probably some kind of nematic order~\cite{chs92,rcc93})
\item and the pyrochlore lattice ~\cite{cl98,mc98}
\footnote{It should be remarked that in the last two cases, the
divergence of Eq.~\ref{reduc} comes from the existence of a
zero mode on the whole Brillouin zone. This zero mode is due
to a local degeneracy of the classical
ground-state~\cite{rcc93}.}
.
\end{itemize}

Table~\ref{ener-_param} summarizes quantitative properties of the
spin-1/2 Heisenberg model on different simple lattices.

\begin{table}
\begin{center}
\begin{tabular}{|c||c|c|c|c|c|c|}
\hline
&Coordination &$2<{\bf S}_i.{\bf S}_j>$	&&\\
Lattices&number&per bond&$M/M_{cl}$&\\
\hline
\hline
dimer &1&-1.5&& \\
\hline
 1 D Chain&2&-0.886&0&\\

honeycomb~\cite{fsl01}&3 &-0.726& 0.44& bipartite\\
sq-hex-dod.~\cite{tr99}&3&-0.721 &0.63&lattices\\
square~\cite{tc90}&4&-0.669& 0.60&\\
\hline
 one triangle&2&-0.5 &&\\
\hline
 kagom\'e~\cite{web98}&4&-0.437&0&frustrating\\
triangular~\cite{bllp94} &6&-0.363&.50&lattices\\
\hline

\end{tabular}
\end{center} 
\caption[99]{ Energy per bond and sublattice magnetization in the
ground-state of the spin-1/2 Heisenberg Hamiltonian on various
lattices. The sq-hex-dod. is a bipartite lattice formed with
squares, hexagons and dodecagons.}
\label{ener-_param}
\end{table}

\subsection{Semi-classical ground-states with competing interactions}
Another way to frustrate N\'eel order and try to destabilize
it, consists in adding competing interactions. The archetype
of such a problem is the $J_1-J_2$ model:
\begin{equation}
{\cal H} = 2 J_1 \sum_{<ij>}  {\bf S}_i. {\bf S}_j + 2 J_2
\sum_{<<ij>>}  {\bf S}_i. {\bf S}_j ,
\label{J_1-J_2}
\end{equation}
where the first sum runs  on first neighbors  and the second on second
neighbors.  As  an example let us  consider  the square  lattice case.
The nearest  neighbor antiferromagnetic  coupling favors a $(\pi,\pi)$
order, which  can be destabilized by  a 2nd neighbor $J_2$ coupling $>
J_{2c}  \sim 0.35$. For  much  larger $J_2$,  the  system recovers some
N\'eel long   range ordering. With a  simple   classical reasoning one
would then expect an order parameter with a 4 sublattice geometry and
an internal degeneracy ~\cite{cd88,ccl90,sz92}\footnote{Notice that the
extra degeneracy is a global one and not a local one as in the kagom\'e
problem.}.  As  a consequence of  quantum  or  thermal fluctuations  a
collinear configuration ($(\pi,0)$ or $(0,\pi)$) is selected  via the
mechanism   of  ``order through    disorder''~\cite{vbcc80,s82}: 
 the collinear  order  is
stabilized by  fluctuations because it is a   state of larger symmetry
with a larger reservoir of  soft fluctuations nearby. In the classical
problem  the stabilization is entropy  driven (whence the  name of the
mechanism), in  the quantum problem  the softer fluctuations  induce a
smaller  zero point   energy and   thus a stabilization   of the  more
symmetric state.   In all  these cases,   the  first excitations are
gapless $\Delta S^z =1$ magnons.

\subsection{One-dimensional problems}
As already  mentioned, the   Heisenberg problem  in that  case  (Bethe
Chain) is critical.    Its low-lying  excitations are  free   spin-1/2
excitations called spinons    (first discovered by   Fadeev).  Due  to
quantum   selection rules   they appear   in  pairs, the   spectrum of
excitation is thus  a continuum with two  soft points ($0$ and $\pi$).
A frustrating  $J_2$ coupling, larger than  the critical  value $J_2^c
\sim 0.2411$, opens a gap in the spectrum and the ground-state becomes
dimerized   (and   two  fold  degenerate)  with   long  range order in
singlet-singlet  correlations.   Majumdar  and Gosh~\cite{mm69}   have
shown that the ground-state of this  problem for $J_2=0.5$ is an exact
product    of   singlet  wave-functions~\cite{c89a}.    The  low-lying
excitations   are   pairs    of   solitons~\cite{ss81},    forming   a
continuum. For specific values   of   the coupling there appears    an
isolated branch of singlets around the point $\pi/2$~\cite{ys97}.
\end{section}

\begin{section}{Valence Bond Crystals}

\subsection{Effect of a frustrating interaction on a N\'eel state}

Coming back to the $J_1-J_2$ model of Eq.~\ref{J_1-J_2}, we examine
the possibility of destroying N\'eel ground-states by a
frustrating interaction.
At the point  of maximum frustration ($J_2 \sim  0.5 J_1$) between the
two   N\'eel phases ($(\pi,\pi)$ and   $(\pi,0)$)  the system can
lower its energy by 
 forming $S=0$ valence bonds between nearest neighbors.
This maximizes the binding energy of the bonds involved in
singlets  and decreases the  frustrating  couplings  between different
singlets.  But  contrary  to   N\'eel states,  nearest   neighbors not
involved in singlets  do not  contribute  to the  stabilization of this
phase: so  the presence of such a  phase between two  different N\'eel
phases  is  not mandatory and  should be  studied  in each  case.  The
nature of the  arrangements of singlets is also  an open question: the
coverings could have long range order or not.  We  call the phases with
long range   order in the  dimer  arrangements:   Valence Bond Crystals
(noted in the following VBC).  The simplest  picture of a Valence Bond
Crystal ground-state is given in  Fig.~\ref{vbc}.  The other kinds  of
dimer covering ground-states will be studied in the next sections.

\begin{figure}
	\begin{center}
	\resizebox{3.5cm}{!}{\includegraphics{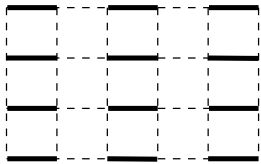}}  \end{center}

	\caption[99]{  A simple  ``columnar''  valence-bond crystal on
	the square lattice: fat links indicate the pair of sites where
	the      spins      are      combined      in   a      singlet
	$\left|\uparrow\downarrow\right>-\left|\downarrow\uparrow\right>$.
	A realistic  VBC   will also  include   fluctuations of  these
	singlet    positions   so   that    the  magnetic  correlation
	$\left<\vec{S}_i\cdot\vec{S}_j\right>$ will    be  larger than
	$-\frac{3}{4}$ along fat links and  will not be zero along the
	dashed ones.}  \label{vbc}
\end{figure}

\subsection{ Ground-state properties of the Valence Bond
Crystals}
\begin{itemize}
\item The ground-state of a VBC is a pure $S=0$ state: it
does not break $SU(2)$ rotational invariance and has only short
range spin-spin correlations, 
\item It has long range dimer-dimer or  $S=0$ plaquettes-plaquettes
correlations~\footnote{A   $S=0$ plaquette is an  ensemble  of an even
number  of nearby  spins arranged in  a  singlet state: a 4-spin S=0
 plaquette long range   order
has been suggested for   the $J_1-J_2$ model on the   square
lattice~\cite{zu96,cs00}, and a 6-spin S=0 plaquette long range   order
 for  the   same  model  on  the hexagonal
lattice~\cite{rs90,fsl01,ms01a}.},
\item In the thermodynamic limit, it breaks the symmetries of
the lattice: translational symmetry and possibly rotational symmetry.
\end{itemize}
Remarks:

Such a    ground-state  is   highly  reminiscent  of     the dimerized
ground-state    appearing  in the $J_1   -J_2$   chain for  $J_2 >
J_2^c$.  But, as explained  below, the  spectrum of  excitations of the
two-dimensional   VBC  is   expected   to   be  different    from  the
one-dimensional case.

On a finite  sample the exact ground-state of  a VBC cannot be reduced
to the symmetry breaking configuration  A of Fig.~\ref{vbc}: it is  the
symmetric superposition of configuration A and of its three transforms
(B, C and D) in  the symmetry operations of the  lattice group. In the
thermodynamic limit, the  4 different superpositions of  the  A, B, C,
and D configurations are  degenerate, allowing the consideration of  a
symmetry   breaking   configuration  like  A     as ``a  thermodynamic
ground-state'' of the   system.  In the general  case  of a   VBC, the
degeneracy of the  thermodynamic ground-state  is finite and  directly
related to  the dimension of the lattice  symmetry group.  On a finite
square sample, the 4  different superpositions of  the A, B, C, and  D
configurations  are non  degenerate~\footnote{ These  4 superpositions
will have  the  following  momenta: $(0,0)$,  $(0,0)$,  $(\pi,0)$  and
$(0,\pi)$.    The   two later  will    be  degenerate (because of  the
$\pi/2$-rotation  symmetry) so that  3 nearly-degenerate energies will
appear in the finite-size spectrum.  }, but their energies are expected
to approach zero exponentially with the system size.

\subsection{Elementary excitations}

Spin-spin correlations are exponentially decreasing and we thus expect
that the  system  will  have a  spin-gap.    The nature of   the first
excitations is not totally settled: it is believed that they should be
$S=1$ bosons  (some kinds  of   optical magnons). The  possibility  of
$S=1/2$ excitations (spinons) is dismissed on the following basis: the
creation of  a pair  of   spins 1/2 by    excitation of a singlet   is
certainly  the basic  mechanism   for  producing an excitation.    The
question  is then:  can these spins  1/2  be separated beyond a finite
distance?  In   doing so they   create a string of  misaligned valence
bonds, the energy of which increases with its length. It is the origin
of an  elastic restoring force  which binds  the two  spins 1/2.  This
simple   picture shows  the influence   of  dimensionality; creating a
default in the $J_1 -J_2$ chain only  affects the neighboring spins of
the chain, the perturbation does not depend on the distance between the
two spins 1/2, in this case the spinons are de-confined giving rise to
a continuum  of  two-particle excitations~\cite{ss81,ys97}. Because of
the  confinement of  spinons in the   VBCs, the excitation spectrum of
this kind of states  will exhibit isolated modes (``optical magnons'')
below the continuum of multi-particle excitations.

An open  question is whether the   effective coupling between  the two
spinons   is preferentially  ferromagnetic  or antiferromagnetic? This
might  be connected to  the preferred position of  the  spinons on the
same (or different) sublattice(s)?  Numerical results on the $J_1-J_2$
model on the hexagonal lattice plead in favor of an antiferromagnetic
coupling between  spinons, manifesting itself by  a  gap in  the $S=0$
sector smaller than in the $S=1$ sector.

As  a consequence of the  gap(s), both  the spin-susceptibility $\chi$
and the specific heat $C_v$ of the VBCs are thermally activated.

\subsection{A toy model}
The simplest 2-dimensional quantum model  exhibiting VBC phases is the
quantum  hard-core dimer  (QHCD)  model   introduced by  Rokshar   and
Kivelson in  the context of high $T_c$ super-conductivity~\cite{rk88}.
It  is  not  strictly speaking  a spin   model since   the Hamiltonian
directly operates on nearest-neighbor dimer  coverings. 
The Hamiltonian  is defined by its leading non-zero matrix
elements within this subspace. On the square lattice it reads:

\begin{equation}
H=\sum_{\rm Plaquette}
\left[
-J\left(
\left|\begin{picture}(13,9)(-2,2)
	\put (0,0) {\line (0,1) {8}}
	\put (8,8) {\line (0,-1) {8}}
	\put (0,0) {\circle*{3}}
	\put (0,8) {\circle*{3}}
	\put (8,0) {\circle*{3}}
	\put (8,8) {\circle*{3}}
	\end{picture}
\right>
\left<\begin{picture}(13,9)(-2,2)
	\put (0,0) {\line (1,0) {8}}
	\put (8,8) {\line (-1,0) {8}}
	\put (0,0) {\circle*{3}}
	\put (0,8) {\circle*{3}}
	\put (8,0) {\circle*{3}}
	\put (8,8) {\circle*{3}}
	\end{picture}
\right|
+h.c.\right)
+V\left(
\left|\begin{picture}(13,9)(-2,2)
	\put (0,0) {\line (0,1) {8}}
	\put (8,8) {\line (0,-1) {8}}
	\put (0,0) {\circle*{3}}
	\put (0,8) {\circle*{3}}
	\put (8,0) {\circle*{3}}
	\put (8,8) {\circle*{3}}
	\end{picture}
\right>
\left<\begin{picture}(13,9)(-2,2)
	\put (0,0) {\line (0,1) {8}}
	\put (8,8) {\line (0,-1) {8}}
	\put (0,0) {\circle*{3}}
	\put (0,8) {\circle*{3}}
	\put (8,0) {\circle*{3}}
	\put (8,8) {\circle*{3}}
	\end{picture}
\right|
+
\left|\begin{picture}(13,9)(-2,2)
	\put (0,0) {\line (1,0) {8}}
	\put (8,8) {\line (-1,0) {8}}
	\put (0,0) {\circle*{3}}
	\put (0,8) {\circle*{3}}
	\put (8,0) {\circle*{3}}
	\put (8,8) {\circle*{3}}
	\end{picture}
\right>
\left<\begin{picture}(13,9)(-2,2)
	\put (0,0) {\line (1,0) {8}}
	\put (8,8) {\line (-1,0) {8}}
	\put (0,0) {\circle*{3}}
	\put (0,8) {\circle*{3}}
	\put (8,0) {\circle*{3}}
	\put (8,8) {\circle*{3}}
	\end{picture}
\right|
\right)
\right]
\label{qhcd}
\end{equation}

The  connection between  this  model  and a spin   model  such as  the
Heisenberg model is in the general case a  difficult issue, because it
involves the  overlap matrix~\footnote{Consider the restriction of the
Heisenberg   Hamiltonian to  the   subspace of nearest-neighbor  dimer
coverings. Due to the non-orthogonality  between dimer coverings, this
effective Hamiltonian is complicated and non-local. However, it can be
formally  expanded  in powers of $x=1/\sqrt{N}$   where $N$ counts the
number of states a spin can have  at each site  (of course in our case
$N=2$).  When  only  the lowest  order ($\mathcal{O}(x^4)$)  terms are
kept, the effective Hamiltonian is given by Eq.~\ref{qhcd}.}  of dimer
configurations   (we     will    come  back    to   this     point  in
section~\ref{section_kag}).  However,  this model can be usefully seen
as  an  effective description of   the low-energy physics of  some non
N\'eel phases.  The  kinetic  term ($J>0$)  favors resonances  between
dimer   configurations which differ  by   two dimers   flips around  a
plaquette.  The potential term ($V>0$) is a repulsion between parallel
dimers.  A   strong negative  $V$   favors   a Valence  Bond  columnar
phase~(Fig.~\ref{vbc}),  whereas  a     strong positive   V    forbids
crystallization  in  the columnar  phase   and favors a staggered  one
(Fig.~\ref{staggered_sq}).  From  our present  point of view,  both of
these phases are gapped VBC.  They are separated by a quantum critical
point  at  $J=V$.  At  that   point  the  exact  ground-state  is  the
equal-amplitude   superposition      of  all   nearest-neighbor  dimer
configurations~\cite{rk88},          the    correlations           are
algebraic~\cite{fs63}, the  gap  closes  and  the  system supports
$S=0$   (quasi-)Goldstone    modes       (called   ``resonons''     in
ref.~\cite{rk88}).

\begin{figure}
	\begin{center}
	\resizebox{3.5cm}{!}{\includegraphics{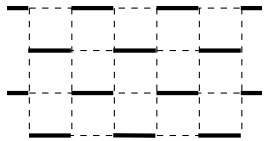}}
	\end{center}
	
	\caption[99]{Staggered dimer configuration stabilized
	by a strong dimer-dimer repulsion $V>0$ in the QHCD on
	the square lattice.	
	}\label{staggered_sq}

\end{figure}

Moessner and Sondhi have recently  studied the same  kind of model  on
the hexagonal lattice: it has a richer  phase diagram of VBC phases,
and essentially the same generic physics~\cite{msc01}.

\subsection{Possible realizations in Spin-1/2 $SU(2)$ models}

The $J_1-J_2$  model on the square  lattice near  the point of maximum
frustration  has    been   studied  by  different   approaches:  exact
diagonalizations,    series   expansions,   Quantum   Monte-Carlo  and
Stochastic Reconfiguration. There   is  a general  agreement  that the
$J_1-J_2$        model     on    the     square~\cite{sz92,zu96,kosw99,cs00},
hexagonal~\cite{fsl01}  and  checker board lattices~\cite{fmsl01} has
at least 
one VBC phase between the two semi-classical N\'eel phases. Long
range order might involve plaquettes of 4 spins (on the square lattice
or on  the checker board  lattice) and even 6   spins on the hexagonal
lattice.

Two elements should be noticed:
\begin{itemize}
\item All these lattices are bipartite and their
coordination number is not extremely large,
\item These phases appear when both $J_1$ and $J_2$ are
antiferromagnetic   (see next  section    for a   ferromagnetic  phase
destabilized by an antiferromagnetic coupling).
\end{itemize}

\subsection{Large-$N$ limits}

Introduced   by Affleck   and  Marston~\cite{am88}  and    Arovas  and
Auerbach~\cite{aa88},  large-$N$  limits    are  powerful  analytical
methods.    The $SU(2)$ algebra  of  a  spin $S$  at   one site can be
represented  by  $N=2$   species of particles   $a^\dag_\sigma$  (with
$\sigma=\uparrow,\downarrow$),   provided   that  the total number  of
particles on one site is constrained to be $a^\dag_\uparrow a_\uparrow
+  a^\dag_\downarrow  a_\downarrow=2S$.  The  raising  operator  $S^+$
(resp. $S^-$) is simply  represented by $a^\dag_\uparrow a_\downarrow$
(resp. $a^\dag_\downarrow a_\uparrow$).  These particles can be chosen
to be fermions (Abrikosov fermions) or bosons (Schwinger bosons).  The
Heisenberg interaction is a four-body interaction for these particles.

The idea of large-$N$ limits is to  generalize the $SU(2)$ symmetry of
the  spin$-S$ algebra to an $SU(N)$   (or $Sp(N)$) symmetry by letting
the  flavors  index $\sigma$  go  from $1$   to $N$.  The $SU(N)$  (or
$Sp(N)$) generalization of the Heisenberg model is  solved by a saddle
point calculation of    the  action, which  decouples   the  different
flavors.  It is equivalent to a mean-field decoupling of the four-body
interaction of the physical $N=2$ model.

Whether  the  large-$N$  limit of some    $N=2$  model is   an accurate
description of  the physics  of  ``real'' $N=2$  spins is a  difficult
question  but the phase  diagrams obtained  using 
these approaches (where
the value  of  the ``spin'' $S$  can be  varied) are  usually coherent
pictures of the competing phases  in the problem. In particular,  these
methods  can  describe  both N\'eel   ordered states and  Valence-Bond
Crystals~\cite{rs90}, as  well as short-range Resonating Valence Bonds
phases (section~\ref{rvb}) or   Valence-Bond Solids~\cite{aklt87}.  Read
and Sachdev~\cite{rs91} studied the  $J_1-J_2-J_3$ model on the square
lattice  by a $Sp(N)$ bosonic representation  and predicted a columnar
VBC phase (as in Fig.~\ref{vbc}) near $J_2\simeq0.5J_1$.

\subsection{Experimental realizations of Valence Bond Crystals}

In  the recent years many  experimental realizations of VBCs have been
studied. The most studied prototype in 1d is $CuGeO$: it is not a pure
realization  of the dimerized phase  of a pure $J_1-J_2$ model insofar
as  a  Spin-Peierls instability   induces a small  alternation  in the
Hamiltonian.  This compound has received a lot  of attention both from
the  experimental and theoretical points of  view (see \cite{ys97} and
references   therein).  In two dimensions  two   compounds are 2D VBCs
CaV$_4$O$_9$~\cite{ki94,ttw94,tnyk95,fo96,am96a,am96b,my96,sr96,uksl96,khst96,tku96,khsk97,oyiu97}
and SrCu$_2$(BO$_3$)$_2$. This last compound is  a good realization of
the Shastry  Sutherland model~\cite{ss81a,k99}.  However, in both case
the ground-state is  non-degenerate  because  the Hamiltonian  has  an
integer spin  in the unit  cell (4 spins  $1/2$)  and the dimerization
does not break any lattice symmetry.

\section{Short-range Resonating Valence-Bond phases: type I SRRVB Spin Liquid}
\label{rvb}

\subsection{Anderson's idea}
Inspired by Pauling's idea of ``resonating  valence bonds'' (RVB) in
metals, P.~W.~Anderson~\cite{a73}  introduced  in  1973 the idea  that
antiferromagnetically coupled spins 1/2 could  have a
ground-state
 completely different from the two previous cases.

An RVB state can be viewed as a linear superposition of an exponential
number of disordered  valence  bond   configurations where spins    are
coupled  by pairs in singlets  (contrary  to the Valence Bond  Crystal
where a finite   number   of  ordered configurations   dominate    the
ground-state wave-function and the physics  of the phase).  Because of
these   singlet pairings, many  configurations  are expected to have a
reasonably low energy  for an antiferromagnetic Hamiltonian.  However,
these energies are not necessarily  lower that the variational  energy
of a  competing N\'eel state.  What lower   the energy of  a RVB state
with respect to the energy of one particular lattice dimerization, are
the resonances between  the exponential number of dimer configurations
which are energetically very close or degenerate. These resonances are
possible because the Hamiltonian (the Heisenberg one for instance) has
non-diagonal  matrix elements between  almost     any pair of    dimer
coverings.

One  central  question  is  to   characterize  the  kinds   of   dimer
configurations  which   have  the  most   important  weights  in   the
wave-function.   The low-energy physics  will  crucially depend on the
separation between the spins which are paired in singlets\footnote{The
set of all dimer   coverings, including singlet  dimers of  {\em all}
lengths,  is in  fact   an over-complete   basis of   the  whole $S=0$
subspace.  Without specifying which  dimer configurations do  enter in
the wave-function, a linear superposition  of dimer configurations can
in fact  be {\em any} $S=0$  state, including a  state with long-range
spin-spin correlations!}.  Both ideas  of short - and long-ranged  RVB
states have been developed in the literature.  Liang {\it et al.} have
shown  that long range order  on  the square lattice  can be recovered
with dimer wave functions as soon as the weight of configurations with
bonds of length l   decreases more slowly  than $l^{-5}$~\cite{lda88}.
In   this  section  we  concentrate  on   the   case where   only {\em
short-ranged}  (i.e.  a  finite number  of  lattice  spacings  but not
necessarily first-neighbors) dimer singlets participate  significantly
in the ground-state wave-function.  In  such a situation, taken  apart
the singular case of  the Valence Bond  Crystals, the short  range RVB
state has no  long range order,  it is  fully invariant under  $SU(2)$
rotations: it  is a genuine  {\em Spin Liquid}~\footnote{ Some authors
use the word spin liquid with a less  restrictive meaning for all spin
systems exhibiting a spin gap inasmuch as they  do not break SU(2). In
view of the  importance of the LRO  in the ground-state  and low lying
excitations of the Valence   Bond Crystals described in the   previous
section, we prefer the present definitions.}.

\subsection{Ground-state properties of type I SRRVB Spin Liquids}
\label{ground-state}

Consider a spin model with a spin-$\frac{1}{2}$ in  the unit cell.  Our
definition of a short-range RVB state is a wave-function which has the
following properties:
\begin{itemize}

\item[(A)] it can be written as a superposition of short-ranged dimer
configurations,

\item[(B)] it has exponentially decaying  spin-spin correlations,
 dimer-dimer correlations and any higher order correlations,

\item[(C)] its ground-state displays a subtle topological
degeneracy~\cite{s88,rk88,rc89,wen91,mlbw99,o00,mlms01}.

\end{itemize}

Let us make some remarks on this definition.  First, property (A) does
not imply (B). As an example, the equal-amplitude superposition of all
nearest-neighbor  dimer    configurations   on   the    {\em   square}
lattice has  algebraic spin-spin correlations~\cite{fs63}. However such
equal-amplitude  superposition  on the   {\em triangular} lattice  was
recently shown to satisfy (B)~\cite{ms01}. Property C is specific to
 systems with half odd integer spins in the crystallographic unit cell.

\subsection{Elementary excitations of type I SRRVB Spin Liquids}

If all kinds of correlations decay  exponentially with distance over a
finite correlation length $\xi$, all  symmetry sectors are expected to
be gapful \footnote{and $\chi$  and $C_v$ are thermally  activated}.
The simplest heuristic     argument  is the    following:   low-energy
excitations are usually obtained by   long wavelength deformations  of
the ground-state order parameter. If the correlation length is finite,
elementary excitations will have  a size of  order $\xi$ (or smaller),
which is the  largest distance in the  problem.  Such excitations will
therefore have a finite energy (uncertainty principle).

\subsection*{Magnons and spinons}

A  conventional  N\'eel  antiferromagnet   has elementary  excitations
called  magnons (or spin-waves)  which  carry an  integer  spin $\Delta
S^z=\pm1$.   In  one   dimension, there are   free  spin-$\frac{1}{2}$
elementary excitations   (spinons).  In two  dimensions,  spinons  are
confined in   VBC phases  but  the  possibility  of unconfined spinons
exists in a short-range RVB phase (naively  speaking there are no more
elastic  forces to  bind spinons  in  disordered dimer  coverings).  A
field-theoretic  description (Large$-N$   limit  and gauge-theory)  of
spinon (de-)confinement  in spin liquids  was carried  out by Read and
Sachdev~\cite{rs91}.  Unconfined spinons are fractional excitations in
the sense that they  carry a  quantum number (total  spin) which  is a
fraction of the local degrees of freedom,  namely the $\Delta S=\pm 1$
spin flips. A  first  example of  a {\em  2D  system} with  unconfined
spin-$\frac{1}{2}$   excitations   might      have     been   observed
experimentally~\cite{coldea00}.  From  the theoretical  point of view,
only    toy  models  have   been  rigorously  demonstrated  to exhibit
spinons~\cite{ms01,ns01}, although  they are a  generic feature in large-$N$
($Sp(N)$) approaches  to  frustrated  spin-liquids  (see  for instance
Refs.~\cite{rs91,cms01,cmm01}) at least at mean-field level.

\subsection{The hard core quantum dimer model on the triangular
lattice}

The  QHCD  model was   originally  introduced to    look for an  RVB
phase. As explained in the previous section,  it turns out that on the
square lattice it  displays only (gapped  or critical) VBC phases.  It
has  been recently generalized  to  the triangular lattice by Moessner
and Sondhi~\cite{ms01}.  Contrary to the original square lattice case,
the QHCD  model on the triangular  lattice  provides a short-range RVB
spin-liquid with a finite   correlation  length at  zero  temperature.
This phase survives  in a finite  interval of parameter $V_c\leq V\leq
J$.  The triangular version of the QHCD model is probably the simplest
microscopic model which exhibits a short-range RVB phase, and
quasi particle deconfinement.

\subsection{Realizations of a type I Spin Liquid in $SU(2)$ spin
models}

Since the  pioneering  work of Anderson,  there  has  been an  intense
theoretical   activity   on  the     RVB  physics,   specially   after
Anderson~\cite{a87} made  the the proposition  that such an insulating
phase could be  closely related to the  mechanism  of high-T$_{\rm c}$
superconductivity.  Short-range RVB states are certainly a stimulating
theoretical concept, their realization in microscopic spin models is a
more complicated issue.  Up to now, two-dimensional models which could
exhibit a short-range RVB phase are still  few: Ising-like models in a
transverse  magnetic field~\cite{msc00} which  are closely  related to
QHCD    models      by    duality~\cite{ns01,msf01},    a    quasi  1d
model~\cite{psz93}, a  spin-orbital model~\cite{ssgpt99},  and the two
short-range RVB phases in $SU(2)$ models, that we will now present.

\subsection*{The multiple-spin exchange model}
The multiple-spin  exchange model  (called  MSE in the  following) was
first introduced by Thouless~\cite{t65}  for the nuclear magnetism  of
three-dimensional         solid    He$^3$~\cite{rhd83}      and     by
Herring~\cite{herring66} 
for  the Wigner crystal.   It is an effective
Hamiltonian which governs the spin degrees of  freedom in a crystal of
fermions.  The Hamiltonian is a sum of permutations which exchange the
spin variables  along rings of  neighboring sites.  It  is now largely
believed that MSE interactions on the triangular lattice also describe
the    magnetism     of  solid   He$^3$    mono-layers    adsorbed  on
graphite~\cite{rbbcg98,mblw98,collin01} and that  it  could be a  good
description   of     the  two   dimensional     Wigner    crystal   of
electrons~\cite{bcc01}. In the He$^3$ system, exchange terms including
up to  6 spins are present~\cite{rbbcg98}. Here  we will only focus on
2- and  4-spin interactions which   constitute the minimal  MSE  model
where  a   short-range   RVB  ground-state is  predicted   from  exact
diagonalizations~\cite{mlbw99}. The Hamiltonian reads:

\begin{equation}
H=J_2 \sum_{
\begin{picture}(17,10)(-2,-2)
	\put (0,0) {\line (1,0) {12}}
	\put (0,0) {\circle*{5}}
	\put (12,0) {\circle*{5}}
\end{picture}
} P_{i j}
+J_4 \sum_{
\begin{picture}(26,15)(-2,-2)
        \put (0,0) {\line (1,0) {12}}
        \put (6,10) {\line (1,0) {12}}
        \put (0,0) {\line (3,5) {6}}
        \put (12,0) {\line (3,5) {6}}
        \put (6,10) {\circle*{5}}
        \put (18,10) {\circle*{5}}
        \put (0,0) {\circle*{5}}
        \put (12,0) {\circle*{5}}
\end{picture}
} \left( P_{i j k l}+P_{l k j i} \right)
\label{J2J4}
\end{equation}

The  first   sum runs  over  all pairs  of  nearest  neighbors  on the
triangular  lattice and $P_{i j}$ exchanges  the spins between the two
sites $i$ and $j$. 
 The second sum runs over  all the 4-sites plaquettes and $P_{i
j k  l}$ is  a cyclic permutation  around the  plaquette.  The  2-spin
exchange  is equivalent to the  Heisenberg  interaction since  $
P_{ij} = 2 \vec{S}_{i}\cdot\vec{S}_{j}  +1/2 $, but the  four-spin term
contains terms involving 2 and 4 spins  and  makes  the model  a highly
frustrated one.

As  $J_2<0$ and $J_2/J_4\simeq -2$ in  low-density solid He$^3$ films,
the   point $J_2=-2$  $J_4=1$   has been studied   by   means of exact
diagonalizations up to  $N=36$ sites~\cite{mlbw99}.   These data point
to a spin-liquid with a  short correlation length  and a spin gap.  No
sign of a  VBC could be  found.  In addition, a topological degeneracy
which characterizes  short-range  RVB states was   observed.  From the
experimental    point  of   view,   early  specific    heat  and  spin
susceptibility  measurements are not  inconsistent with  a spin-liquid
phase  in  solid   He$^3$   films.   Indeed  very  low     temperature
magnetization   measurements~\cite{collin01}   suggest  a    small but
non-zero spin gap in low-density mono-layers.

An explanation of the origin of this short-range RVB  phase in the MSE
model can be guessed  from the analogy between multiple-spin  interactions
and QHCD models.  From the analysis of QHCD  models we understand that
RVB phases are possible when VBC are energetically unstable.  Columnar
VBC are   stabilized  by   strong    parallel dimer  attraction     and
staggered~\cite{ms01} VBC appear   when  the repulsion  between  these
parallel    dimers  is strong.    In    between,  an   RVB  phase  can
arise~\footnote{In  Ref.~\cite{ns01}   an   RVB state  is   selected by
introducing defects in the   lattice   in order to destabilize     the
competing VBC  states.}.    From this   point  of  view,  tuning   the
dimer-dimer interactions is  of  great  importance and the    four-spin
interaction of model (\ref{J2J4}) plays this role~\cite{mlbw99}.

\subsection*{A type I SRRVB phase on the hexagonal lattice}

A second RVB phase has recently been found on the hexagonal lattice in
a    highly frustrating regime where    the  three first neighbors are
coupled ferro-magnetically whereas the  six second neighbor  couplings
are antiferromagnetic~\cite{fsl01}.  For weak second neighbor coupling,
the  ground-state is a ferromagnet.    Increasing the second  neighbor
coupling leads to an instability toward a short range RVB phase which
has all the above-mentioned     properties (except  the    topological
degeneracy, because there are two spins 1/2 in the unit cell).
This kind of phase has possibly been  observed years ago by
L.P. Regnault and J. Rossat-Mignod in $BaCo_2(AsO_4)_2$\cite{rrm90}.

These two RVB phases appear in the  vicinity of a ferromagnetic phase:
is it or is it not an essential ingredient to form an  RVB phase?  One might
argue that this feature helps in favoring RVB phases against VBC ones,
because the plausible VBCs would  have large elementary plaquettes and
would  be very sensitive to resonances  between the different forms of
plaquettes, thus  disrupting  long range  order.   It  should  also be
noticed that the first    neighbor coupling being  ferromagnetic,  the
short range RVB phase will predominantly form on second neighbors, and
thus on a triangular lattice. So the properties of the dimer coverings
on the  triangular  lattice  might at the end be  the  essential
ingredient to have a RVB phase!

\subsection{Chiral spin-liquid}

The  definition  of a   short-range    RVB spin-liquid  proposed    in
section~\ref{ground-state} excludes any long  range order.  However, a
state with  broken {\em time-reversal}  symmetry and chiral long range
order could accommodate all the other properties of a spin-liquid (the
chiral    observable   is the   triple  product  of  three  spins: see
ref.~\cite{wwz89} for various  equivalent definitions).  Such a chiral
phase would have a doubly degenerate ground-state in the thermodynamic
limit.   Inspired  from  Laughlin's   fractional   quantum Hall   wave
functions,   Kalmeyer   and   Laughlin~\cite{kl87,kl89}  have  build a
spin-$\frac{1}{2}$ state on the triangular  lattice which exhibit some
chiral long-range order  (see also~\cite{ywg93}).   This {\em complex}
wave  function is   directly     obtained  from the  bosonic     $m=2$
($\nu=\frac{1}{m}$) Laughlin wave-function.   Such a  state is a  spin
singlet   with  unconfined   spinon   excitations which  have  anyonic
statistics (right in between Bose and  Fermi).  This chiral liquid was
initially    proposed       for    the triangular-lattice   Heisenberg
antiferromagnet  but the later turned    out to be N\'eel   long-range
ordered~\cite{bllp94}.  Wen   {\it  et al}~\cite{wwz89}  discussed the
properties expected for a chiral spin-liquid, its excitations and some
possible  mean-field  descriptions.  To  our   knowledge, there  is no
example  of a  microscopic $P-$  and   $T-$symmetric spin  model which
exhibits  a chiral spin-liquid phase.  The  possibility of realizing a
chiral  phase  in the  presence of an   external magnetic field (which
explicitly   breaks  the  time-reversal  invariance)  is  discussed in
paragraph~\ref{chernsimons}.
\end{section}

\section{Type II SRRVB phases: ``the kagom\'e -like magnets''}
\label{section_kag}

On triangular    based lattices,  de-stabilization   of  the  coplanar
3-sublattice N\'eel order {\it either  by an  increase of the  quantum
fluctuations} (through a   decrease of  the coordination number   when
going  from  the triangular lattice  to the  kagom\'e one)  {\it or by
adding competing interactions} (4-spin exchange processes) leads to an
unexpected situation where the degeneracy of the exponential number of
short range dimer  coverings   is only marginally lifted    by quantum
resonances,   giving  rise to a  quantum  system  with a  continuum of
singlet excitations adjacent to the ground-state. This property has first been
shown by exact  diagonalizations of the  Heisenberg Hamiltonian on the
kagom\'e lattice~\cite{web98}, then  for   the MSE Hamiltonian  on  the
triangular lattice~\cite{lmsl00}.

\subsection{Description of the ground-state and of the first excitations in
the $S=0$ sector}

The  ground-state is a trivial superposition  of an exponential number
of singlets, like  in any RVB ground-state  described in  the previous
section.  But contrary  to the situations   described in the  previous
sections there is no gap above the ground-state in the singlet sector.

Mambrini    and  Mila~\cite{mm01}  have   shown   that the  qualitative
properties of the ground-state and   the  first excitations are   well
described in the restricted   basis of nearest neighbor couplings:  to
this extent, this  second spin-liquid is  a real short-range RVB  state
(indeed dressing these   states with longer dimer  coverings  improves
quantitatively the energy, but does not change the picture).

To understand the   mechanism  of (non)formation   of the  gap  in the
kagom\'e spin-liquid, it is  interesting to compare Mambrini's results
to   the earlier work  of Zeng  and Elser~\cite{ze90}. This comparison
shows that the  non orthogonality of the  dimer basis is  an essential
ingredient  to  produce the  continuum of  singlets   adjacent  to the
ground-state.   The   above-mentioned QHCD     model  which implicitly
truncates the expansion   in the overlaps   of dimers is  by the  fact
unable to describe such a phase. On the other hand, taking into account
the non orthogonality of dimer   configurations would generate a  QHCD
model involving an    infinite expansion   of  n-dimers  kinetic   and
potential terms. In  this basis  the effective Hamiltonian  describing
the original Heisenberg problem has an infinite range of
exponentially decreasing matrix elements.

This system has a $T=0$ residual entropy  in the singlet sector ($\sim
ln(1.15)$ per spin)~\cite{web98,m98}.  The $S=0$ excitations cannot be
described as Goldstone  modes of a  quasi  long range order in  dimers
(similar to the  critical point of  the R.K. QHCD model of  subsection
2.4).  In  such a  system,  the density  of  states would  increase as
$\exp( N^{\alpha/(\alpha+D)})$, where D  is the dimensionality of  the
lattice,  and $\alpha$ the power  index of the  dispersion  law of the
excitations ($\epsilon (k)   = |k|^\alpha$), whereas  it  increases as
$\exp( N)$  in the present   case  (this represents  a large  numerical
difference~\cite{lsf01}).

\subsection{Excitations in the $S \neq 0$ sectors}

The magnetic  excitations  are probably gapped:  this assumption  is a
weak  one.  The spin  gap   if it exists  is  small  (of the order  of
$J/20$).

In each $S \neq 0$ sector the density of low lying excitations 
increase exponentially with the system size as the S=0 density of
states, but with extra prefactors (as for example a $N$
prefactor in the case of the $S =1/2$
sector~\cite{lblps97,web98,m98,mm01}).

The elementary  excitations  are de-confined  spinons~\cite{web98}.   An
excitation in the $S=1/2$ sector  could be seen  as a dressed spin-1/2
in  the sea  of   dimers~\cite{m98}. The  picture  of   Uemura  {\it et
al}~\cite{ukkll94} drawn  from the analysis of  muon data on $SrCrGaO$
is   perfectly  supported  by   exact  diagonalizations  results.  The
analytic description of such  a phase  remains a
challenge. The fermionic $SU(N)$ description~\cite{mz91a,sg01} might give a
good point of departure: here 
the  $S=0$ ground-state is indeed degenerate  in
the saddle point approximation. But the  difficulty of the analysis of
this  degenerate  ground-state in   a  $1/N$ expansion  remains  to be
solved! A recent attempt to deal with such problems in
a dynamical mean field approach looks promising~\cite{gsf01}.

\subsection{Experimental realizations}
No   perfect   $S=1/2$ kagom\'e antiferromagnet    has  been up to now
synthesized.

   An   organic    composite   $S=1$ system    has  been
studied experimentally~\cite{wkyoya97}: it displays a large spin gap ( of the order of the supposed-to-be coupling constant and thus
much larger than what
is   expected  on the   basis of   the  $S=1/2$ calculations).   It is
difficult to claim that it is an experimental manifestation of an
  even-odd integer effect, because the
ferromagnetic binding of the spins 1/2 in a spin 1 is not so large that
the identities  of the underlying compounds could  not play a role. 
(From a theoretical point of view, it would be extremely
interesting to have exact spectra of a spin-1 kagom\'e
antiferromagnet: if topological effects are essential to the
physics of the spin-1/2 kagom\'e one might expect completely
different spectra for the spin-1 system.)

A spin-$3/2$   bilayer    of  kagom\'e planes, the $SrCrGa$  oxide
    has  been  extensively
studied~\cite{r00}.  It  displays some  features that  could readily be
explained in the  present  framework of  the spin-1/2  theoretical
model:
\begin{itemize}
\item Dynamics of the low lying magnetic excitations~\cite{ukkll94}
\item Vanishing elastic scattering at low temperature~\cite{lbar96}
\item Very low sensitivity of the low -T specific heat to very large
magnetic fields~\cite{rhw00,smlbpwe00}
\end{itemize}
but    some   features    (essentially   the  anomalous     spin glass
behavior~\cite{kklllwutdg96,wdvhc00})  remain   to  be  explained in  a
consistent way.

There are also a large  number of magnetic  compounds with a pyrochlore
 lattice (corner   sharing tetrahedra). At the  classical  level
such  Heisenberg magnets have   ground-states with a larger degeneracy
than the  kagom\'e  problem~\cite{mc98}.   They are  expected  to  give
spin-liquids~\cite{cl98}, and indeed some   of them display   no frozen
magnetization~\cite{hztws94}. Whether  the  Heisenberg nearest neighbor
problem for spin-1/2 has the  same generic properties
on the pyrochlore lattice
and on  the kagom\'e lattice is still  an open question. Contrary to
some   expectations~\cite{pc01},   the   Heisenberg   problem   on  the
checker-board  lattice (a  2-dimensional     pyrochlore) has  a    VBC
ground-state~\cite{fmsl01,mts01}. Nevertheless
   it  is  up  to  now  totally
unclear if the checker-board  problem is a  correct description of the
3d pyrochlores, there are even small indications that this could
be untrue~\cite{fmsl01}.

\section{Magnetization processes}
\label{section_mag}

In this section we discuss some aspects  of the behavior of Heisenberg
spin systems in the presence of an external magnetic field. Frustrated
magnets either with a semi-classical ground-state or in purely quantum
phases exhibit a large number of specific magnetic behaviors:
metamagnetism, magnetization plateaus. 
In the first subsection we describe the free energy patterns
associated with these various behaviors. In subsection 5.~2, we
discuss a classical and a quantum criterion for the appearance of
a magnetization plateau. In the following subsections we
described two mechanisms recently proposed to explain the
formation of a plateau.

\subsection{Magnetization curves and free energy patterns}

The simplest quantum antiferromagnet in  2D is the  spin$-\frac{1}{2}$
Heisenberg model on the square lattice:
\begin{equation}
	H=J \sum_{\left<i,j \right>} \vec{S}_i\cdot\vec{S}_j
	- B \sum_i S_i^z
\end{equation}

\begin{figure}
	\begin{center}
	\resizebox{!}{2cm}{\includegraphics{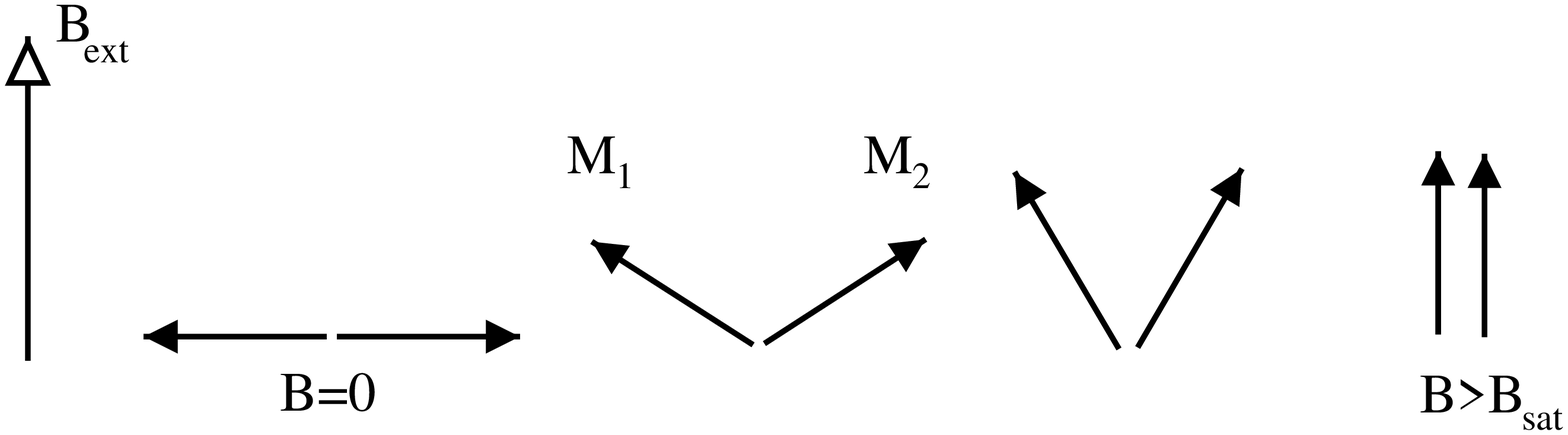}}  \end{center}
	\caption[99]{Schematic view of the  sublattice magnetization vectors when
	the external magnetic  field is increased in an two-sublattice
	Heisenberg antiferromagnet.} \label{mag_sq}
\end{figure}

The  system has two sublattices with   opposite magnetizations in zero
field and  at zero temperature.   The full magnetization curve of this
model  has  been   obtained by    numerical  and analytical
approaches~\cite{zn98}.     The    sublattice magnetizations gradually
rotates toward  the applied field direction  as the magnetic  field is
increased (Fig.~\ref{mag_sq}).  At some  finite critical field $B_{\rm
sat}$ the total  magnetization reaches saturation\footnote{As in  most
models, the saturation field can be computed  exactly by comparing the
energy of the  ferromagnetic state
($E_F/N=J/2-B/2$) 
with  the  (exact) energy of  a  ferromagnetic
magnon ($E({\bf k}\ne{\bf 0})=E_F+B+J(\cos(k_x)+\cos(k_y)-2)$).    The
magnon energy  is   minimum in ${\bf   k}_{\rm  min}=(\pi,\pi)$.   The
saturation field is obtained when  $E_F=E({\bf k}_{\rm min})$, that is
$B_{\rm sat}=4J$. The calculation is  unchanged for an arbitrary value
of the spin $S$ and one finds $B_{\rm sat}=8JS$.}.

\begin{figure}
	\begin{center}
	\resizebox{6cm}{!}{\includegraphics{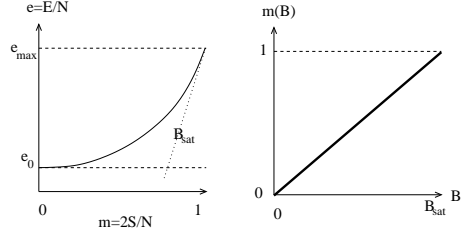}}
	\end{center}
	\caption[99]{Linear response obtained with   $e\simeq
	\frac{m^2}{2\chi}$ as in an AF system with collinear LRO.}
	\label{embA}
\end{figure}

We  define $e(m)$ as the  energy per site $e=E/N$ of  the system, as a
function of the net magnetization $m=\frac{M}{\Msat}$.  From this zero
field  information,  we get   the full  magnetization curve $m(B)$  by
minimizing  $e(m)-mB$,  that is   $B=\dedm$.   In  the  square-lattice
antiferromagnet case discussed    above, $e(m)$ is   almost  quadratic
$e(m)=\frac{m^2}{2\chi_0}$ and   the corresponding magnetization    is
almost linear (Fig.~\ref{embA}). 

When the  system is  more  complicated,  because  of frustration   for
instance, the magnetization process  can be more complex. In particular,
magnetization plateaus   or metamagnetic  transitions  can occur.  For
instance,   the addition  of   a  second-neighbor   coupling on    the
square-lattice antiferromagnet  opens  a plateau at  one   half of the
saturated  magnetization~\cite{h00}.  It is useful to translate
these anomalies of the magnetization curve into properties of $e(m)$.

\begin{figure}
	\begin{center}
	\resizebox{6cm}{!}{\includegraphics{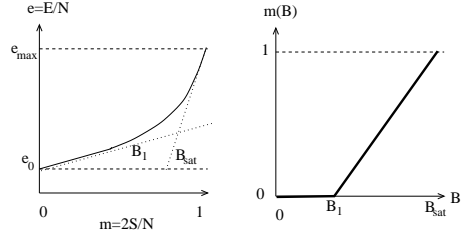}}
	\end{center}
	\caption[99]{$m=0$  magnetization  plateau  due  to   a linear
	$e(m)=  \Delta  m+o(m)$. } \label{embB}
\end{figure}

A    plateau at  $m_0=0$   (also called  spin  gap)   is equivalent to
$\dedm_{|m_0=0}  > 0$ (Fig.~\ref{embB}).  At the   field where the magnetization  starts
growing the transition can be  critical or  first  order.  In one
dimension, exact results  on the Bose condensation of
a dilute     gas   of  interacting magnons    lead to 
$m\sim\sqrt{\delta B}$ for integer spin  chains~\cite{a91}.  In  that
case  $e(m)=\Delta m + am^3 +o(m^3)$ and the system is critical
at $B=\Delta$.

\begin{figure}
	\begin{center}
	\resizebox{6cm}{!}{\includegraphics{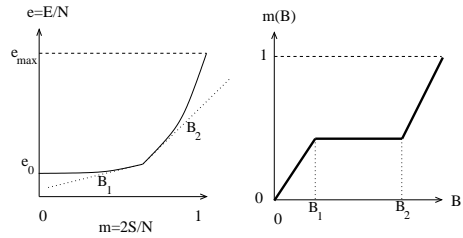}}
	\end{center}
	\caption[99]{A magnetization    plateau  originates  from    a
	discontinuity in the slope of $e(m)$.  The simplest example of
	quantum  Heisenberg   model   with  such   a   plateau is  the
	triangular-lattice antiferromagnet (see Fig.~\ref{magtri}  and
	text), a more sophisticated example is the behavior of
the same model on a kagom\'e lattice, which equally displays a
magnetization plateau at $M/M_{sat}=1/3$.  }  \label{embD}
\end{figure}

A plateau at  finite  magnetization $m_0>0$ for  $B\in[B_1,B_2]$  is a
discontinuity   of     $\dedm$   (Fig.~\ref{embD}). Such a behavior
can arise in a frustrated one
 (see next subsection).   The   vanishing
susceptibility when $B$ is inside the plateau comes from the fact that
magnetic   excitations  (which  increase   or    decrease   the  total
magnetization) are gapped when the magnetic field lies in the interval
$[B_1,B_2]$.  As before,  the system can  be critical at the ``edges''
of the plateaus. In the  following subsections 
we examine the origins of such
plateaus in classical and quantum spin systems.

\begin{figure}
	\begin{center}
	\resizebox{6cm}{!}{\includegraphics{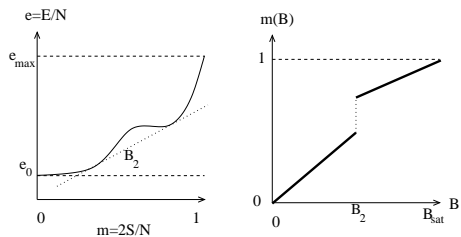}}
	\end{center}
	\caption[99]{Metamagnetic transition  associated
	to a  concavity    in  the   $e(m)$   curve.}
	\label{embC}
\end{figure}

A metamagnetic transition is a discontinuity of the magnetization as a
function  of the applied field, it  is a  first order phase transition
and is  equivalent to a  concavity in $e(m)$ (Fig.~\ref{embC}).   As in
any  first  order  transition, this  can   give rise to an  hysteretic
behavior in experiments. Such a behavior is highly probable in
frustrated magnets~\cite{mblw98,rrm90} for essentially two
reasons:
\begin{itemize}
\item Due to the frustration, different configurations of spins 
corresponding to phases with different space symmetry breakings
are very near in energy,
\item Impurities in magnetic compounds pin the existing
structures and hinder the first order phase transitions, at
variance with their role in the standard liquid-gas transition.
\end{itemize}

As a last  general, but disconnected,  remark let us underline 
that the magnetic studies of quantum
frustrated magnets
 could also help in solving
elusive questions relative to the existence of an exotic H=0
ground-state.
We have already  seen that the extraordinary
thermo-magnetic behavior of $SrCrGaO$ could be a signature of a
type II Spin Liquid~\cite{rhw00,smlbpwe00}.
 Years ago L.P. Regnault and J. Rossat-Mignod studied
$BaCo_2(AsO_4)_2$: the knowledge of its magnetic phase diagram
convinced them that something queer was going on in this material
at $H=0$ and in fact exact diagonalizations now point to a type
I Spin Liquid!
In fact in very
large magnetic fields the  magnets become increasingly classical
and a semi-classical approach  is justified: any
deviation from a semi-classical behavior when H is decreasing
 is thus an important indication of a possible quantum exotic ground-state! 

\subsection{Magnetization plateaus}

\subsection*{Classical spins: collinearity criterion}
\label{collinearity}

\begin{figure}
	\begin{center}
	\resizebox{2.5cm}{!}{\includegraphics{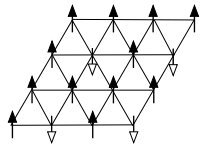}}
	\end{center}
	
	\caption[99]{Collinear  spin structure of type  $uuud$ on
	the triangular lattice.  This state with $M/\Msat=\frac{1}{2}$
	is realized   in the  classical  and  quantum MSE  model under
	magnetic  field      (see   text    and   Fig.~\ref{magMSE}).}
	
	\label{uuud}
\end{figure}

Magnetization   plateaus  are    often believed   to   be  a    purely
quantum-mechanical phenomenon   which  is sometimes  compared  in  the
literature  to Haldane phases of integer  spin  chains.  This is
certainly not always true since  some {\em classical} spin models have
magnetization plateaus at zero temperature.  For instance, as shown by
Kubo and  Momoi~\cite{km97}, the MSE  model on  the triangular lattice
has    a large  range of  parameters  where   magnetization plateaus  at
$M/\Msat=\frac{1}{3}$   and   $M/\Msat=\frac{1}{2}$  appear at   zero
temperature~\footnote{As  discussed  below,   these plateaus  are also
present           in          the         quantum      $S=\frac{1}{2}$
model~\cite{mblw98,msk99,lmsl00}.}.  The ground-state of the system at
$M/\Msat=\frac{1}{3}$ is the so-called $uud$ structure where two
sublattices have  spins pointing ``up''  along the field  axis and the
third one has  down spins.  At $M/\Msat=\frac{1}{2}$ the  ground-state
is of $uuud$ type (Fig.~\ref{uuud}).

\begin{figure}
	\begin{center}
	\resizebox{6cm}{!}{\includegraphics{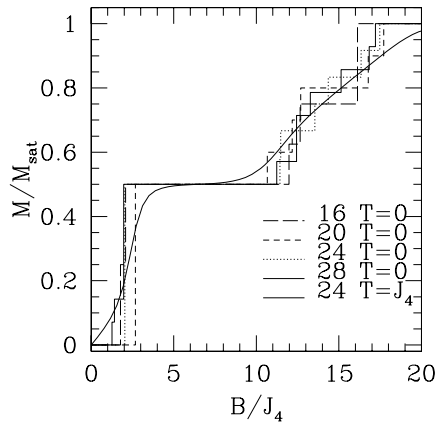}}
	
	\end{center} \caption[99]{Magnetization curve of the MSE model
	at $J_2=-2$  and $J_4=1$ computed numerically~\cite{msk99,m99}
	for finite-size   samples with $N$  up   to  $28$ sites.   The
	magnetization plateau  at  $M/\Msat=\frac{1}{2}$  (due to  the
	$uuud$ state  of   Fig.~\ref{uuud}) clearly  appears  even  at
	finite temperature (full line~\cite{m99}).}

	\label{magMSE}
\end{figure}

In fact, we show   below that if  a classical  spin system  exhibits a
magnetization plateau then all the spins  are necessarily  collinear
to the magnetic field.  This  restricts possible spin configurations
to states of  the $u^{n-p}d^{p}$ kind with $n-p$  spins ``up'' and $p$
spins  ``down''  in   the  unit   cell.   As a   consequence,    the total
magnetization per site must be of the form: $M/\Msat=1-2p/n$
where $n$  is the size of  the unit cell, and $p$ an integer.
  This commensuration between
the magnetization value  at a plateau and  the total spin in  the unit
cell will turn out to hold equally for quantum spins.

The proof of this classical  collinearity condition is sketched below.
We consider  the function $e(m)$   (which is defined without  external
field).  Let us  suppose that the ground state  $\Psi_0$ at $m_0$ is {\em
not}  a  collinear configuration.  We   will  show that $e(m)$  has  a
continuous  derivative in  $m_0$, {\em  i.e.}    no plateau  occurs at
$m_0$.  From a non-collinear state, one can chose an angle $\theta$ to
deform  $\Psi_0$ into   a  new  configuration  $\Psi(\theta)$   with a
different magnetization
\begin{equation}
	m(\theta)=m_0+a\theta+{\mathcal{O}}(\theta^2) \label{m}
\end{equation}  and $a\neq0$. This can
be done, for instance,  by rotating the  spins  of a  sublattice whose
magnetization  is not collinear to  the  field  \footnote{If $\Psi_0$
was  a  collinear  state,  the  deviation  of magnetization   would be
smaller: $m(\theta)=m_0+b\theta^2+{\mathcal{O}}(\theta^3)$   and   the
argument  would fail.}.  Here we  just  require that the  spins can  be
rotated   of  an   infinitesimal  angle,  as  it   is  the   case  for
three-component classical spins.  $\Psi(\theta)$ is a priori no longer
the         ground      state         and       its       energy    is:
\begin{equation}
e(\Psi(\theta))=e(m_0)+\alpha\theta+{\mathcal{O}}(\theta^2)
\label{e_Psi}
\end{equation}
No  assumption  on $\alpha$  is needed.   As  a  variational state  of
magnetization    $m(\theta)$,  $\Psi(\theta)$       has   an    energy
$e(\Psi(\theta))$ larger or equal to  the  the ground state energy  at
magnetization $m(\theta)$:
\begin{equation}
	e(m_0)+\alpha\theta+{\mathcal{O}}(\theta^2) \geq e(m(\theta))
\end{equation}
With Eq.~\ref{m} we see that:
\begin{equation}
	e(m_0)+\frac{\alpha}{a} \left(m(\theta)-m_0\right)
	+{\mathcal{O}}\left((m(\theta)-m_0)^2\right) \label{le} \geq  e(m(\theta))
\end{equation}
If    we  assume that no     metamagnetic   transition occurs in   the
neighborhood  of  $m_0$,  then   $e(m)$  is convex  about    $m_0$ and
Eq.~\ref{le} insures the differentiability  of $e(m)$ in $m=m_0$.   We have
$\dedm_{|m=m_0^-}=\dedm_{|m=m_0^+}$ and no plateau occurs at $m_0$.

This  proof is not restricted to  rotation invariant interactions.  We
have only  made the (extremely weak) assumption  that the energy  is a
continuous    and  differentiable function    of  the spins directions
(Eq.~\ref{e_Psi}).       If    the        couplings    are     written
$J^x\vec{S}^x_i\vec{S}^x_j                  +J^y\vec{S}^y_i\vec{S}^y_j
+J^z\vec{S}^z_i\vec{S}^z_j$, a plateau must  correspond to a collinear
state  with  respect to  the  magnetic  field  direction, whatever the
easy-plane or easy-axis might be.   This property remains also true if
multiple-spin interactions are present.

\subsection*{Fluctuations}

Zero-temperature magnetization plateaus will  of course be smeared out
at very    high   temperature.   However,   in  some   cases,  thermal
fluctuations can enhance magnetizations plateaus.  In  the case of the
triangular-lattice       classical  Heisenberg antiferromagnet     the
magnetization is perfectly  linear  at zero temperature for  isotropic
Heisenberg interactions.  It is  thanks to thermal fluctuations that a
$uud$ structure is stabilized   and that the susceptibility  is reduced
(but  not  zero) about  $M/\Msat\simeq\frac{1}{3}$~\cite{km85}  (quasi
plateau).  This phenomenon is due to the  higher symmetry of collinear
states compared with  other nearby spin configurations.  This symmetry
enlarges the available phase space  volume for low-energy fluctuations
and selects the $uud$ phase.

Quantum  fluctuations  can  play a  very   similar role:  the  quantum
$S=\frac{1}{2}$  Heisenberg model on the  triangular lattice has
 an exact
magnetization      plateau    at   $M/\Msat=\frac{1}{3}$     at   zero
temperature~\cite{nm86,cg91,h99}.  Again, the ground-state  at
$M/\Msat=\frac{1}{3}$ is a $uud$-like state.  This $uud$ plateau has a
simple       origin   in   the  Ising    limit~\cite{nm86,h99}
($H=J^z\sum_{\left<i,j\right>}S^z_iS^z_j$)) but  survives  up   to the
isotropic point  $J^z=J^{xy}$.  What  should  be stressed  is that the
introduction of quantum fluctuations through   couplings in the   $xy$
plane  renormalizes    the  magnetic   field   interval    where   the
susceptibility vanishes but  the total  magnetization at  the  plateau
remains {\em exactly} $\frac{1}{3}$.  The sublattice magnetizations of
the two $u$ sublattices and the $d$  sublattice are reduced from their
classical  values to $M_u=1-\eta$ and  $M_d=-(1-2\eta)$  but the total
magnetization is unaffected $M=(2M_u+M_d)=\frac{1}{3}\Msat$.  The same
phenomenon was observed numerically for the $uuud$  plateau of the the
MSE model on the triangular lattice which also exists in the classical
limit~\cite{km97} and      survives quantum  fluctuations     in   the
spin$-\frac{1}{2}$ model~\cite{mblw98,msk99,lmsl00} without  change in
the magnetization value.  In the following paragraphs we discuss the origin
of this robustness of the magnetization value at a plateau.

\begin{figure}
	\begin{center}
	\resizebox{!}{2cm}{\includegraphics{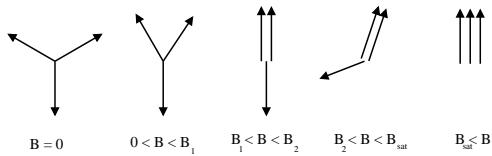}}
	\end{center}
	\caption{ Magnetization  process of    the  spin$-\frac{1}{2}$
	Heisenberg antiferromagnet  on  the triangular  lattice.   The
	magnetic field   is along the  vertical  and the three vectors
	represent   the   three   sublattice    magnetizations.     A
	plateau at $M/\Msat=\frac{1}{3}$ is present for
	magnetic      field    between          $B_1$              and
	$B_2$~\cite{nm86,cg91,h99}.}  \label{magtri}
\end{figure}

\subsection*{Quantum spins and Oshikawa's criterion}

The Lieb-Schultz-Mattis (LSM)  theorem~\cite{lsm61} proves  that in one
dimension and in the   absence of magnetic  field the  ground-state is
either   degenerate or   gapless  if the   spin  $S$ at  each  site is
half-integer.   The proof relies on  the  construction of a low energy
variational state which is orthogonal to  the ground state.  Oshikawa,
Yamanaka and Affleck~\cite{oya97}  realized   that this proof  can  be
readily  extended to magnetized  states as long  as $n(S-s^z)$ is {\em
not} an integer, where $s^z$ is  the magnetization per site  and $n$
the  period of the ground  state~\footnote{$n$ can be larger than what
is    prescribed  by the      Hamiltonian   if the ground-state   {\em
spontaneously} breaks  the translation symmetry.}.  As a magnetization
plateau requires a   spin gap, this   suggests that plateaus  can only
occur  when $n(S-s^z)$ is an  integer.   However the low-energy  state
produced by LSM has the same magnetization $s^z$  as the ground state,
- it  is a non-magnetic  excitation-, and  does not  exclude a gap to
{\em magnetic} excitations when $n(S-s^z)$ is not  a integer.  In fact
arguments based  on  the  bosonization  technique were    provided by
Oshikawa   {\it et  al.}~\cite{oya97}  (see   also Ref.~\cite{t98})  to
support  the  hypothesis  that  $n(S-s^z)\in \mathbb{Z}$  is   indeed a
necessary  condition to have    a magnetization plateau
 in one-dimensional systems.   Since  then
numerous   studies  of    magnetization   plateaus in  one-dimensional
systems~\cite{chp97,h99a,hmt00} confirmed  that all the plateaus
satisfy $n(S-s^z)\in \mathbb{Z}$.    We   will see   below  that   this
criterion appears also to apply in higher dimension.

The LSM construction  does not  give a  low energy state  in dimension
higher than one  but Oshikawa~\cite{o00} developed a  different
approach to relate the magnetization  value to the  number of spins in 
 the unit
cell.   His  result applies  to itinerant particles on 
 lattice models  when   the  number of
particles is a conserved quantum number.  It states that a gap is only
possible when the  number of particle in  the unit cell is an integer.
Since spins  $S$ are  exactly  represented by  interacting bosons, his
results       applies    to   quantum    magnets    and   reduces   to
$n(S-s^z)\in \mathbb{Z}$.  The  hypothesis  that  one  has to   make to
establish   the result  is  the  following: the   system has  periodic
boundary conditions  (it is a $d-$dimensional torus)  and  the gap, if
any,  does not close when adiabatically  inserting one fictitious flux
quantum inside    the  torus~\footnote{The  number  of   sites  in the
$(d-1)$-dimensional section of the torus is supposed to  be odd.}.  In
the  spin language,  this  amounts to say that  the  spin gap does not
close    when   twisting   the  boundary    conditions   from   0   to
$2\pi$~\cite{mlms01}.  This    hypothesis    can  only   be    checked
numerically~\cite{mlms01}  but it is  completely  consistent with  the
idea that the   gapped system is a  liquid  with a finite  correlation
length   and  which  physical properties  do  not   depend on boundary
conditions in the thermodynamic limit.

All known    examples  of magnetization plateaus  indeed    do satisfy
Oshikawa's  criterion.  It    is interesting  to   notice  the   close
resemblance between the requirement that  $n(S-s^z)$ is an integer and
the  collinearity condition discussed in paragraph~\ref{collinearity}.
Let us assume that we  can represent one spin $S$ with  $2S$
classical  spins of length $\frac{1}{2}$ and  require that these spins
are in a  collinear configuration.  If the unit  cell has $n$ sites it
contains  $n'=2Sn$ small spins,   $p$  of which     are   down
($u^{n'-p}d^p$).  For one unit cell the magnetization is $M=n'/2-p$ so
that the magnetization  per site  is  $s^z=M/n=S-p/n$.  The  classical
criterion reads $n(s^z-S)=-p\in\mathbb{Z}$ which is the same condition
as    Oshikawa's  result.  In spite   of   this  striking analogy, the
classical  collinearity  condition is  an  exact  result  valid in any
dimension and it does not involve  a topological property (as
periodic boundary conditions on the sample) unlike
the  quantum version. Ultimately indeed it rests on the property
that the discretization of a spin S only involves spin-1/2
units.

\subsection{Magnetization plateaus as crystal of magnetic particles}

Magnetization plateaus can often be  understood simply
if the system has some
unit cell where  spins are strongly coupled  (exchange $J_1$) and weak
bonds $J_2 << J_1$  between  the cells.   In this case,  magnetization
plateaus are governed by the quantized magnetization of a single cell,
and the plateaus can be  continuously connected to the un-coupled cells
limit $J_2=0$.  A pertubative  calculation in powers of $J_2/J_1$ will
in general predict the plateau to be stable over  a finite interval of
$J_2$. This situation appears, for instance, in one-dimensional ladder
systems~\cite{h99a}.     In   2D,    an    example      is   the
$M/\Msat=\frac{1}{2}$       plateau  predicted~\cite{fo99}    in   the
$1/5$-depleted square lattice  realized in CaV$_4$O$_9$.  This lattice
is  made of coupled   square plaquettes and  the $M/\Msat=\frac{1}{2}$
plateau comes from the  magnetization   curve of a  single   plaquette
(which  magnetization    can of   course   take  only   three  values:
$M/\Msat=0,\frac{1}{2}$ and $1$).

However, in a translation invariant  2D system, with  a single spin in
the unit cell, the mechanism for the  appearance of plateaus is not so
simple.   The  $M/\Msat=\frac{1}{3}$ plateau of the triangular-lattice
Heisenberg antiferromagnet      cannot  be  understood    in   such  a
strong-coupling picture.   The   magnetization plateaus  predicted  at
magnetizations smaller than  $\frac{1}{2}\Msat$ in the  $1/5$-depleted
square lattice mentioned above cannot either be understood within such
a picture.

Totsuka~\cite{t98} and  Momoi and Totsuka~\cite{mt00}  have associated
magnetization  plateaus  with   the  crystallization   of   ``magnetic
particles''.  In their picture the zero-field ground-state is a vacuum
which is populated  by  bosonic particles  when the  external magnetic
field  is  turned on.    Depending  on the   model and its  zero-field
ground-state these bosons can  represent different microscopic degrees
of freedom: a spin flip on a  single site, an $S^z=1$  state on a link
or a magnetic state of a larger number of spins.  These bosons obey an
hard-core constraint  and carry one quantum  of  magnetization so that
the  magnetization   of the  system  is  proportional  to the particle
density.   The   magnetic field couples  to  the  bosons as a chemical
potential and the boson compressibility is the magnetic susceptibility
of  the spins.   As for  the  antiferromagnetic interactions  (such as
$J\vec{S}_i\cdot\vec{S}_j$),  they   generate   kinetic  as   well  as
repulsive interactions  terms  for these bosons.  On  general grounds,
Momoi and Totsuka expect this gas of interacting bosons to be either a
Bose condensate (superfluid)   or a crystal (or charge-density  wave).
On one hand,  the finite compressibility of  the superfluid leads to a
continuously  varying  magnetization    as  a  function  of   magnetic
field~\footnote{In  some    cases,  the  superfluid     component   is
believed~\cite{mt00}  to   coexist   with  a   crystal  phase  between
magnetization    plateaus (supersolid).}.   On   the  other  hand, the
underlying   lattice will  make  the  crystal  incompressible (density
fluctuations   are gapped)    which   precisely  corresponds    to   a
magnetization plateau.  Such a crystalline arrangement of the magnetic
moments  is also consistent  with the quasi-classical picture provided
by $u^{n-p}d^p$-like states.

In  this approach, the densities ({\it   i.e} magnetizations) at which
the bosons crystallize ({\it i.e}  form plateaus) is mainly determined
by the  range and  strength   of the  boson-boson  repulsion  and  the
geometry of  the lattice.  In some  toy models where the kinetic terms
for  the    bosons vanishes   (hopping is    forbidden by  the lattice
geometry), this allows  to   demonstrate rigorously the existence   of
magnetization     plateaus~\cite{msku00}.   Looking   at   the spatial
structures which minimizes  the repulsion (and thus neglecting kinetic
terms in  the boson Hamiltonian)   gives useful hints of  the possible
plateaus.  Since  these  structures are  stabilized by  the  repulsive
interactions, they can be  compared to a  Wigner crystal  (except that
particles are bosons, not fermions).

\begin{figure}
	\begin{center}
	\resizebox{3.cm}{!}{\includegraphics{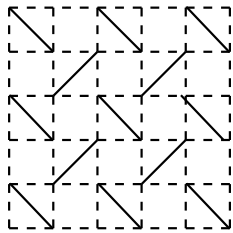}}
	\end{center}  \caption{Shastry-Sutherland lattice. Full   lines
	correspond to Heisenberg exchange $J$ and  dotted ones to $J'$.
	When  $0<J'<0.7J$  the ground-state is    an exact product  of
	singlet states along all  the $J$ links.  This ground state is
	very likely to  be  realized in SrCu$_2$(BO$_3$)$_2$  (in zero
	field) where the couplings   are evaluated to  be  $J'/J\simeq
	0.68$.~\cite{mu99}}  \label{ShaSu}
\end{figure}

The magnetization curve of the quasi 2D oxide SrCu$_2$(BO$_3$)$_2$ has
been  measured    at     very   low    temperatures    up    to     57
Tesla~\cite{k99,okn00},      which    corresponds        to
$M/\Msat\simeq\frac{1}{3}$.  The  magnetization curve displays a large
spin gap and plateaus at $\frac{1}{4}$  and $\frac{1}{3}$. A small one
at $\frac{1}{8}$ is    also  reported.  This  compound   is the  first
experimental  realization of the Shastry-Sutherland  model~\cite{ss81a}
(see Fig.~\ref{ShaSu}).   In  this system magnetic  excitations have a
very small dispersion.  This can  be understood  from the geometry  of
this particular    lattice~\cite{mu99,mu00} and has   been
confirmed    experimentally      by   inelastic  neutron    scattering
experiments~\cite{k00}.  This  low  kinetic energy  makes
 the bosons quasi-localized  objects and    explains   their   ability  to
crystallize~\cite{mu00,mt00b} and give magnetization plateaus.

\subsection{More exotic states - analogy with quantum Hall effect}
\label{chernsimons}

A   recent piece of  work   on   magnetization  plateaus  in   two-dimensional
spin-$\frac{1}{2}$       magnets   by    Misguich,  Jolic{\oe}ur   and
Girvin~\cite{mjg01}  establishes  a connection between this phenomenon
and quantized plateau of the (integer) quantum  Hall effect.  Although
both   phenomena show  up as  quantized  plateaus in  a 2D interacting
system, they    are apparently not directly    related~: magnetization
plateaus involve spins  on a lattice  whereas the quantum  Hall effect
appears  with  fermions  in  the continuum   with   plateaus in  their
transverse  Hall conductance $\sigma_{xy}$.  The  link between the two
problems appears  when the spins are  represented as fermions attached
to one quantum of fictitious magnetic flux, as explained below.

A spin-$\frac{1}{2}$     model    is    equivalent   to      hard-core
bosons~\footnote{Raising  and lowering  operators  $S^+_i$ and $S^-_i$
are equivalent to bosonic creation  and annihilation operators $b^+_i$
and  $b_i$   for    bosons    satisfying  an     hard-core   constraint
$b^+_ib_i=S^z_i+\frac 1 2\leq 1$.}.   It  is possible to  map  exactly
these  bosons to a  model of fermions  in the presence of a fictitious
gauge field~\footnote{  In one dimension, this   mapping from spins to
fermions is   the famous  Jordan-Wigner  transformation and  no  extra
degree of freedom is required.  It is a bit more  involved in 2D where
introducing a fictitious magnetic  field is necessary.}.  The  idea is
that the hard-core  constraint will be  automatically  satisfied by the
Pauli   principle  and the  fictitious flux   quantum attached to each
fermion will    transmute   the  Fermi   statistics  into   a  bosonic
one~\cite{fr89,es92,lrf94}.

In this  framework a down  spin is an  empty site and an  up spin is a
composite object made of one (spinless) fermion and  one vortex in the
fictitious  gauge field centered   on a  neighboring plaquette.   This
vortex can be simply pictured as an infinitely thin soleno\"{\i}d piercing
the plane through a plaquette adjacent to the site of the fermion.  As
two  of these fermion+flux   objects are exchanged  adiabatically, the
$-1$ factor due to the  Pauli principle is  exactly compensated by the
factor $-1$  of the Aharonov-Bohm  effect of one  charge making half a
turn around  a flux quantum.  Consequently,   these objects are bosons
and represent faithfully the spin-$\frac{1}{2}$ algebra.  Technically,
the flux attachment is performed by adding  a Chern-Simons term in the
Lagrangian of the  model, the role of which is to enforce the constraint that
each fictitious flux is tied to one fermion.

At this stage  the spin problem is  formulated as fermions interacting
with a Chern-Simons gauge theory.  A mean field approximation which is
not possible in the original  spin formulation is now transparent: the
gauge  field     can     be   replaced    by      its   static    mean
value~\cite{fr89,ywg93}.   Since  this  static  flux $\Phi$   per
plaquette   comes from  the flux   tubes   initially attached to  each
fermion, it      is    proportional    to  the      fermion    density
$\Phi=2\pi\left<c^+c\right>$.  Because of this  flux, each energy band
splits into sub-bands with a complicated structure.  When this magnetic
field is spatially uniform the band structure  has a fractal structure
as  a  function   of  the   flux    which   is called    a  Hofstadter
``butterfly''~\cite{h76}.    The   mean-field ground-state is
obtained by filling these lowest  energy sub-bands with fermions until
their   density satisfies $\Phi=2\pi\left<c^+c\right>$.   For  a given
flux,  one can then  integrate over the  density of states  to get the
ground-state energy. This  energy as a  function  of the flux  (or the
fermion     density)     is      equivalent       to  $e(m)$     since
$2m+\frac{1}{2}=\left<c^+c\right>$   and   one   can     compute   the
magnetization curve.  Magnetization plateaus open when some particular
band-crossings appear in the Hofstadter spectrum~\cite{mjg01}.

This approximation scheme  was applied to the Shastry-Sutherland model
(Fig.~\ref{ShaSu}) and  a good quantitative agreement~\cite{mjg01} was
found    with  the    experimental   results of      Onizuka {\it   et
al.}~\cite{okn00} on  SrCu$_2$(BO$_3$)$_2$.  In  particular,   the
magnetization plateaus at $M/\Msat=0$, $\frac{1}{4}$ and $\frac{1}{3}$
were reproduced.  The $M/\Msat=\frac{1}{3}$  due to the $uud$ state on
the triangular-lattice antiferromagnet  has also  been described  with
this technique~\cite{mjg01}.

This method is indeed very similar to the Chern-Simons approach to the
fractional quantum Hall  effect~\cite{zhang92} in  which real fermions
(electrons)   are represented  as  hard-core bosons  carrying $m$ (odd
integer) flux quanta~\footnote{These bosons interact with a fictitious
magnetic field  as  well as  with the real  magnetic  field.  In  the
mean-field approximation  mentioned   above the fictitious   field is
replaced by its static  average.   The   real  and averaged    fictitious
magnetic field  exactly  cancel when the  Landau level  filling factor
$\nu$ is $\nu=\frac{1}{m}$.  When this  cancelation occurs the bosons
no longer feel  any magnetic field  and they can Bose-condense.  A gap
opens  and gives  rise  to the  fractional quantum  Hall effect.}.  In
particular, as in the quantum Hall effect,  the system is characterized
by a quantized response coefficient.  In the  quantum Hall effect this
quantity is the transverse conductance $\sigma_{xy}$ which relates the
electric  field to the  charge current  in the perpendicular direction
and in the spin system it relates the spin current with a Zeeman field
gradient in the perpendicular direction~\cite{ha95}. In the mean field
approximation this transverse spin   conductance is obtained from  the
TKNN integers~\cite{tknn82} of the associated Hofstadter spectrum.

As in the quantum Hall effect, the topological nature of the quantized
conductance        protects  the         plateaus       from  Gaussian
fluctuations~\cite{ywg93,mjg01} of the (fictitious) gauge field around
its mean-field value.  If this  mean-field theory captures the physics
of magnetization plateaus,  this topological picture of  the quantized
magnetization establishes a  deep   connection with   the  conductance
plateaus in the Quantum Hall Effect.

There could not be now any definitive conclusion, neither on a set
of conditions both necessary and sufficient to have a plateau in
two-dimensional frustrated systems, nor on the existence of other
mechanisms than those presented in sections 5.3 and 5.4. But
magnetization plateaus as well as metamagnetic transitions are
probably rather ubiquitous properties of frustrated magnets which
still deserve both experimental and theoretical studies. 

{\bf Acknowledgments}: We thank Chitra and R. Moessner for their
careful reading of the manuscript and their valuable suggestions.

\end{document}